\documentstyle[12pt,epsfig,axodraw]{article}

\def\baselinestretch{1.1}
\advance\textheight by \topskip
\begin{document}
\tolerance=100000
\thispagestyle{empty}
\setcounter{page}{0}

\newcommand{\be}{\begin{equation}}
\newcommand{\ee}{\end{equation}}
\newcommand{\br}{\begin{eqnarray}}
\newcommand{\er}{\end{eqnarray}}
\newcommand{\ba}{\begin{array}}
\newcommand{\ea}{\end{array}}
\newcommand{\bi}{\begin{itemize}}
\newcommand{\ei}{\end{itemize}}
\newcommand{\bn}{\begin{enumerate}}
\newcommand{\en}{\end{enumerate}}
\newcommand{\bc}{\begin{center}}
\newcommand{\ec}{\end{center}}
\newcommand{\ul}{\underline}
\newcommand{\ol}{\overline}
\def\epem{\ifmmode{e^+ e^-} \else{$e^+ e^-$} \fi}
\def\mathrm{\rm}
\newcommand{\WW}{$W^+ W^-$}
\newcommand{\eeww}{$e^+e^-\rightarrow W^+ W^-$}
\newcommand{\LEPONE}{LEP1}
\newcommand{\LEPTWO}{LEP2}
\newcommand{\QCD}{QCD}
\newcommand{\JETSET}{JETSET}
\newcommand{\MC}{MC}
\newcommand{\qqgg}{$q\bar q gg$}
\newcommand{\qqQQ}{$q\bar q Q\bar Q$}
\newcommand{\eeWWqqQQ}{$e^+e^-\rightarrow W^+ W^-\ar q\bar q Q\bar Q$}
\newcommand{\eeqqgg}{$e^+e^-\rightarrow q\bar q gg$}
\newcommand{\eeqqQQ}{$e^+e^-\rightarrow q\bar q Q\bar Q$}
\newcommand{\eewwjjjj}{$e^+e^-\rightarrow W^+ W^-\rightarrow 4~{\rm{jet}}$}
\newcommand{\eeqqggjjjj}{$e^+e^-\rightarrow q\bar 
q gg\rightarrow 4~{\rm{jet}}$}
\newcommand{\eeqqQQjjjj}{$e^+e^-\rightarrow q\bar q Q\bar Q\rightarrow
4~{\rm{jet}}$}
\newcommand{\eejjjj}{$e^+e^-\rightarrow 4~{\rm{jet}}$}
\newcommand{\jjjj}{$4~{\rm{jet}}$}
\newcommand{\qbar}{$\bar q$}
\newcommand{\Qbar}{$\bar Q$}
\newcommand{\ar}{\rightarrow}
\newcommand{\sm}{${\cal {SM}}$}
\newcommand{\Dir}{\kern -6.4pt\Big{/}}
\newcommand{\Dirin}{\kern -10.4pt\Big{/}\kern 4.4pt}
\newcommand{\DDir}{\kern -7.6pt\Big{/}}
\newcommand{\DGir}{\kern -6.0pt\Big{/}}
\newcommand{\wwqqqq}{$W^+ W^-\ar q\bar q Q\bar Q$}

\def\Ord{\buildrel{\scriptscriptstyle <}\over{\scriptscriptstyle\sim}}
\def\OOrd{\buildrel{\scriptscriptstyle >}\over{\scriptscriptstyle\sim}}
\def\pl #1 #2 #3 {{\it Phys.~Lett.} {\bf#1} (#2) #3}
\def\np #1 #2 #3 {{\it Nucl.~Phys.} {\bf#1} (#2) #3}
\def\zp #1 #2 #3 {{\it Z.~Phys.} {\bf#1} (#2) #3}
\def\jp #1 #2 #3 {{\it J.~Phys.} {\bf#1} (#2) #3}
\def\pr #1 #2 #3 {{\it Phys.~Rev.} {\bf#1} (#2) #3}
\def\prep #1 #2 #3 {{\it Phys.~Rep.} {\bf#1} (#2) #3}
\def\prl #1 #2 #3 {{\it Phys.~Rev.~Lett.} {\bf#1} (#2) #3}
\def\mpl #1 #2 #3 {{\it Mod.~Phys.~Lett.} {\bf#1} (#2) #3}
\def\rmp #1 #2 #3 {{\it Rev. Mod. Phys.} {\bf#1} (#2) #3}
\def\cpc #1 #2 #3 {{\it Comp. Phys. Commun.} {\bf#1} (#2) #3}
\def\sjnp #1 #2 #3 {{\it Sov. J. Nucl. Phys.} {\bf#1} (#2) #3}
\def\xx #1 #2 #3 {{\bf#1}, (#2) #3}
\def\preprint{{\it preprint}}
\def\hepph #1 {{\tt hep-ph/#1}}

\begin{flushleft}
{RAL-TR-2000-039}
\end{flushleft}
\begin{flushright}
\vskip-1.0truecm
{LC-TH-2000-047}\\
{August 2000}
\end{flushright}
\vskip0.1cm\noindent
\begin{center}
{\Large {\bf The implementation of the four-jet
matrix elements\\
in HERWIG and elsewhere\footnote{Talk given at the 
2nd ECFA/DESY Study on Physics 
and Detectors for a Linear Electron-Positron Collider, 
Padova, Italy,  4-8 May 2000.}}}\\[0.5cm]
{\large 
S.~Moretti}\\[0.05 cm]
{\it Rutherford Appleton Laboratory, Chilton, Didcot, Oxon OX11 0QX, UK.}
\end{center}

\begin{abstract}
{\small
\noindent
The new version \cite{herwig61} of the HERWIG event generator 
\cite{HERWIG} allows one to
simulate hadronic final states in electron-positron annihilations
using the ${\cal O}(\alpha_s^2)$ partonic
matrix elements (MEs) for $e^+e^-\to q\bar qgg$ 
and $e^+e^-\to q\bar q Q\bar Q$. 
The consequent phenomenology is discussed in the case of four-jet signatures 
of particular relevance to future electron-positron Linear Colliders (LCs),
such as in Higgs boson searches.
The new  approach adopted in HERWIG is compared to the
standard ${\cal O}(\alpha_s)$ description, as well as to other recently
presented event generators also based on $e^+e^-\to4$~parton MEs. 
}
\end{abstract}
\noindent
Four-jet events will play a crucial role at future LCs \cite{ee500}. 
Just one example is sufficient to make this point. Such machines
are an ideal laboratory where to pursue studies in the Higgs sector
\cite{HiggsNLC}.
In this respect, it should be recalled that
$e^+e^-\to 4~\mathrm{jet}$ processes from QCD will represent a 
serious noise in the search for Higgs particles, both in the standard 
electroweak (EW) theory
and in possible extensions. In particular, in the Standard Model (SM), 
the dominant Higgs production channel proceeds  
via the `bremsstrahlung' process $e^+e^-\ar HZ$ \cite{Bjorken}, 
followed by the hadronic decay $H\ar b\bar b$ (in the intermediate
$M_H$ regime), 
with the $Z$ boson going into pairs of jets about 70\% of the times.

Thus,  it is very important that four-jet events from QCD   are correctly
implemented in the Monte Carlo (MC) programs that are widely used in 
phenomenological studies of hadron production at $\epem$ colliders. 
In this connection,  it has always been rather worrying that certain 
aspects of four-jet production were never  well described by the 
{standard} 
`${\cal O}(\alpha_s)$ ME + parton shower (PS)' MC programs. As an 
illustration, one may recall the
four-jet studies performed by the ALEPH Collaboration at LEP1 
\cite{ALEPHgluino}, which had revealed a significant disagreement
between data and MCs in the typical four-jet angular variables
(see Ref.~\cite{ioebas} for the definition and the discussion of
some of their properties):
(i) the Bengtsson-Zerwas angle $\chi_{\mathrm{BZ}}$; 
(ii) the K\"orner-Schierholz-Willrodt angle $\Phi_{\mathrm{KSW}}$;
(iii) the (modified) Nachtmann-Reiter angle $\theta_{\mathrm{NR}}^*$;
(iv) the angle between the two least energetic jets $\theta_{34}$.

\begin{figure}[!t]
~\epsfig{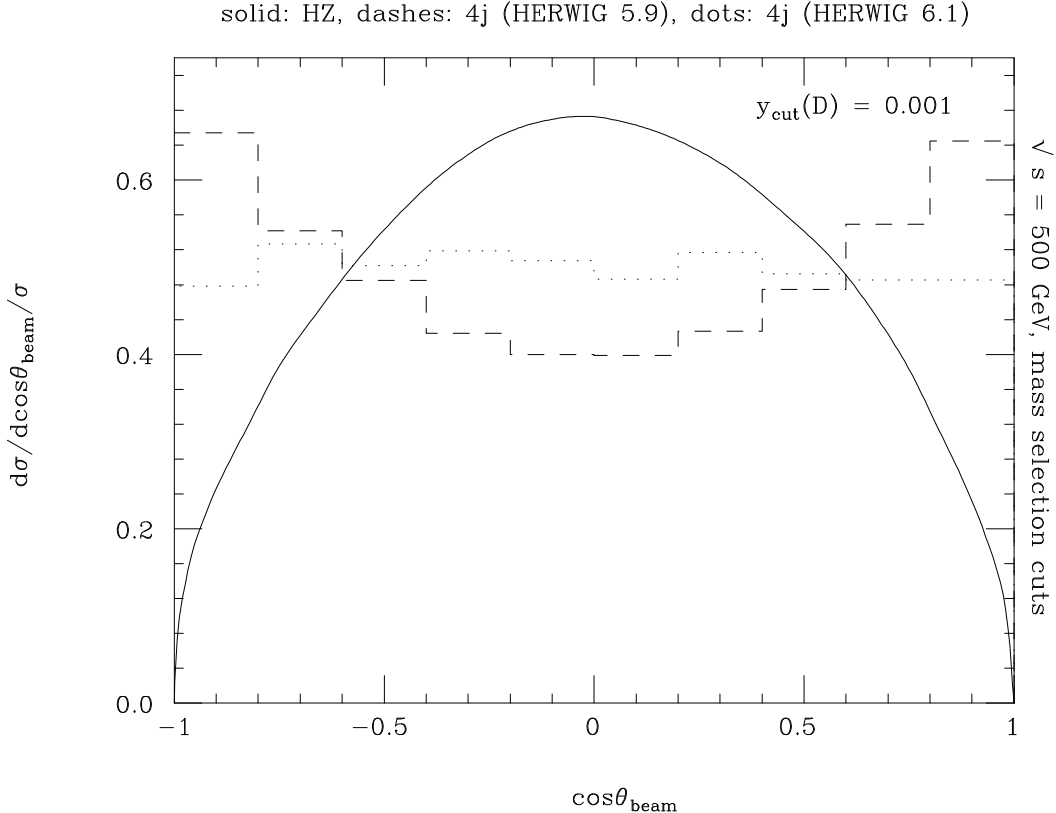}
\caption{\small Differential distributions for the polar angle
of the reconstructed $Z$.}
\label{fig:polar}
\end{figure}

In Ref.~\cite{wrksp}, it was  argued that the discrepancy was nothing but the
evidence that the standard 
`${\cal O}(\alpha_s)$ ME $+~{\mathrm{PS}}$'  MCs do not
provide a correct description of the spin correlations among the various 
partons, particularly at large jet separations. In fact, the
above angular quantities are defined 
in terms of the three-momenta of the
particles in the final state and are precisely the spins of the latter 
that induce very peculiar orientations in (i)--(iv).  
Conversely, `${\cal O}(\alpha_s^2)$ ME' programs
(for example, the  `${\cal O}(\alpha_s^2)$ ME 
+ $\mbox{string fragmentation}$ model' as implemented in JETSET, i.e.
with no parton shower) yielded a much
better angular description of four-jet final states, see 
Refs.~\cite{ALEPHgluino,GC}.
In fact, all spin  correlations are naturally included in a full
${\cal O}(\alpha_s^2)$ ME calculation, 
but they are not necessarily present  in a 
PS emulation based on the ${\cal O}(\alpha_s)$ scattering, 
although their availability
(in the infrared limit, where soft and collinear correlations can
be factorised in analytic form) is a feature of some of the 
above mentioned MC codes (for example, for the HERWIG implementation, 
see Ref.~\cite{Ian}). However, even an `${\cal O}(\alpha_s^2)$ ME 
+ $\mbox{fragmentation}$' model, without the PS evolution,
is inadequate to describe QCD four-jet production in $e^+e^-$
collisions. The problem is  that  
such ME models contain `ad-hoc' hadronisation which  is adjusted
to produce a good agreement with some data  \cite{ALEPHgluino,OP},
but they cannot be 
extrapolated to other energies. Furthermore,  their description of the 
sub-jet structure is very poor (see Ref.~\cite{Andre} for an overview).

That such deficiencies in the description of four-jet final states
could be cured by an `${\cal O}(\alpha_s^2)$ ME 
$+~{\mathrm{PS}}$' (plus hadronisation) approach has been clear for some time.
Now, several event generators have been made available,
that resort to this implementation: HERWIG 6.1 \cite{herwig61},
PYTHIA 6.1 \cite{pythia61}, 4JPHACT \cite{4jphact} and APACIC++
\cite{apacic}.
It is the intent of this note to describe the impact that the
new description of four-jet final states
can have in some contexts which are particularly
crucial to the physics of future LCs. To do so, we
will borrow as an example the case of the HERWIG event generator,
though many of its features can equally be appreciated in the case of
the other MCs. Eventually, we will also highlight the differences and 
similarities among these various programs. 

\begin{figure}[!t]
~\epsfig{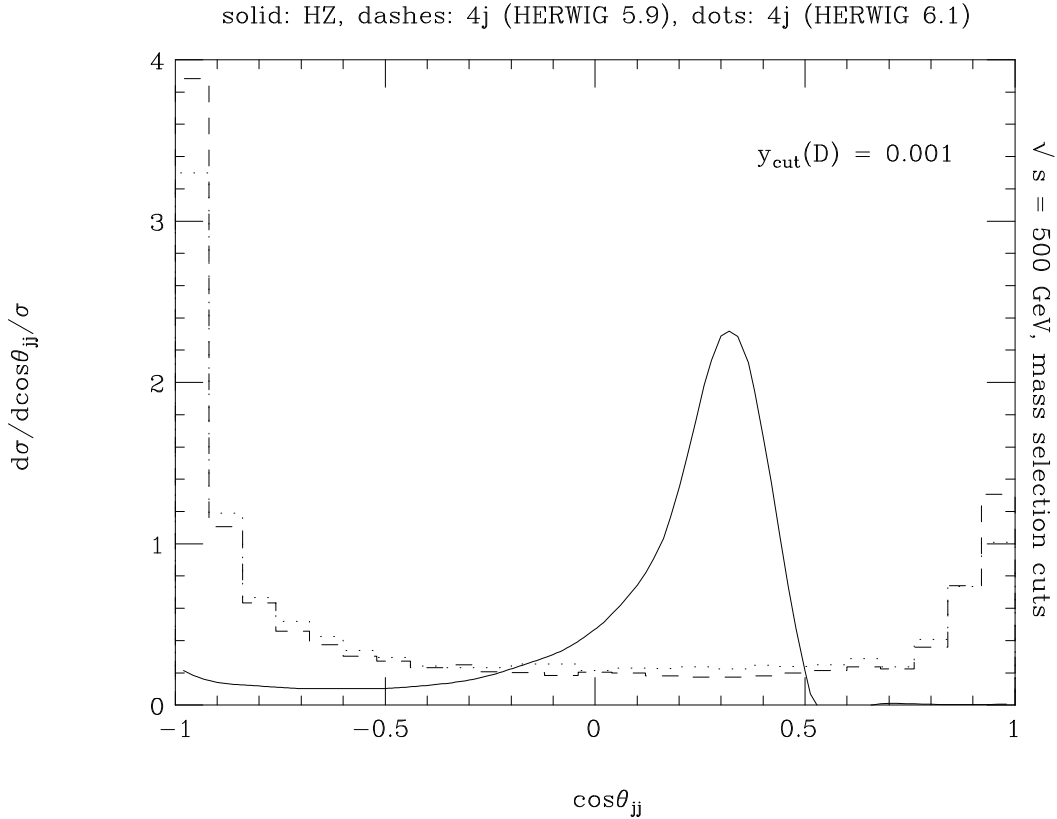}
\caption{\small Differential distributions for the opening angle of the jets 
in  the di-jet system reconstructing the $Z$.}
\label{fig:cjj}
\end{figure}

Ref.~\cite{spin} has already made the point, for the case of LEP,
that systematic differences exist between the use of the 
${\cal O}(\alpha_s)$ and ${\cal O}(\alpha_s^2)$ MEs in the simulation
of QCD four-jet events. This is true not only for the
case of QCD studies (e.g., in the determination of the
QCD colour factors) but also in the context of $M_{W^\pm}$ measurements.
Here, we illustrate similar effects taking place in a higher energy 
environment, such as a LC operating at $\sqrt s=500$ GeV. 

Before proceeding to the numerical part of our
note, we would like to spend a few words about some technical aspects
of the HERWIG implementation. The new 
process describing electron-positron annihilation into four jets
can be simulated by setting {\tt IPROC=600+IQ}, where a non-zero value for 
{\tt IQ}
guarantees production of quark flavour {\tt IQ} whereas
 {\tt IQ=0} corresponds to
the natural flavour mix. ({\tt IPROC=650+IQ} is as above but without those
terms in the MEs which orient the event with respect
to (w.r.t) the lepton
beam direction.) The MEs are based on those of Ref.~\cite{ERT}
 with orientation terms from \cite{CataniSeymour}. Infrared 
(soft and collinear) divergences are avoided  by imposing a minimum separation
among the four partons 
(hereafter, denoted by $y_{\mathrm{inf}}$ --   {\tt Y4JT} in the code, 
by default equal to $0.01$). At present,
the jet separation is calculated
using either the Durham or Jade metrics 
(see, e.g. Ref.~\cite{schemes}, for definitions and a review 
on jet clustering algorithms).  This choice is governed by
the logical variable {\tt DURHAM} (whose default is {\tt .TRUE.}). 
Note that the phase space is for massless partons, as are the MEs,
though a mass threshold cut is applied. The 
 scale {\tt EMSCA} for the  parton showers is set equal to
$\sqrt{s y_{\mathrm{min}}}$,
where $y_{\mathrm{min}}$  is the
least  distance, according to the selected metric,
between any two partons. Finally, the jet resolution $y_{\mathrm{cut}}$ -- used
to select a multi-jet final state -- should
be chosen such that it is always larger than $y_{\mathrm{inf}}$.   

\begin{figure}[!t]
~\epsfig{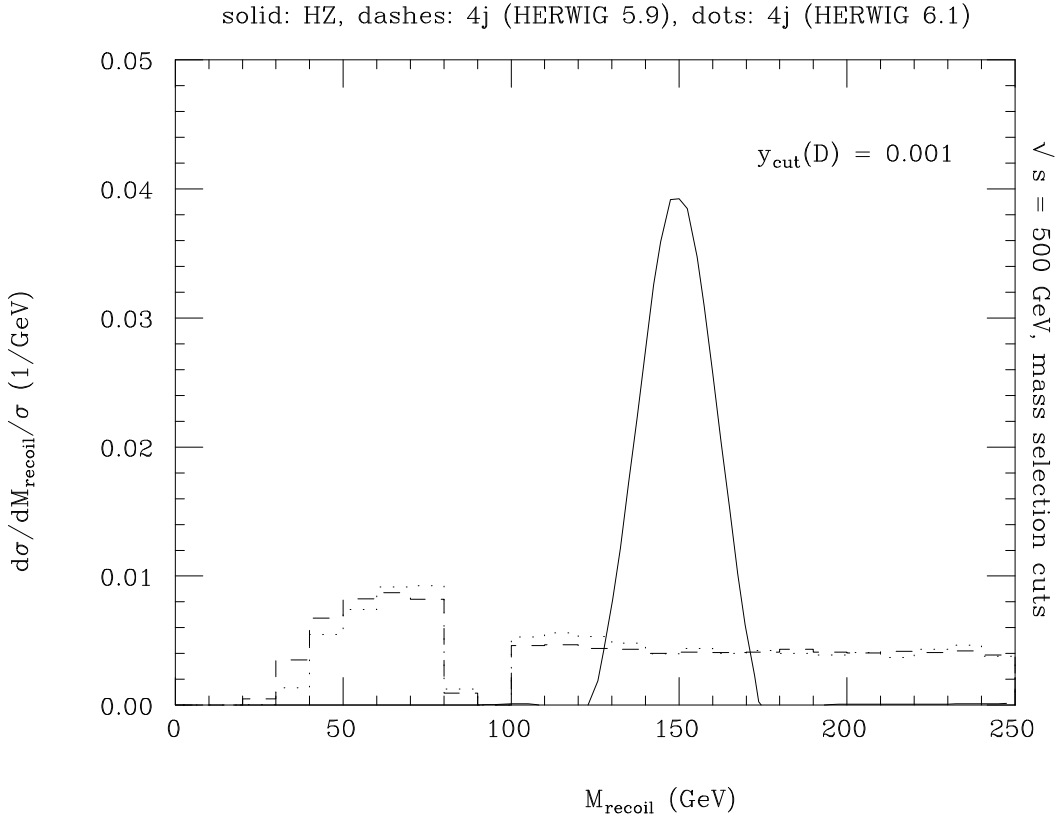}
\caption{\small Differential distributions for the invariant mass recoiling
against the reconstructed $Z$.}
\label{fig:MH}
\end{figure}

An ambiguity  arises in the treatment of interference terms
between diagrams with different colour structures. In fact, apart from
the trivial case $e^+e^-\to q\bar q Q\bar Q$, with $q\ne Q$, the MEs 
for $e^+e^-\to q\bar qgg$ 
and $e^+e^-\to q\bar q q\bar q$ (i.e. $q=Q$)
receive contributions from two distinct colour flows each.
In general, the interference terms between the latter
are not positive definite and a choice has to be made, on how to assign them 
to one or the other of the two leading colour contributions 
(although suppressed 
by $1/N_C^2$ w.r.t. the latter, they cannot be neglected without
inducing sizable effects in both total and differential rates).
For illustration, we describe here the treatment adopted in HERWIG for 
 $e^+e^-\to q\bar q gg$ and refer the reader to the forthcoming new
HERWIG manual \cite{herwig62}
(see also \cite{kosuke}) for details on the
$e^+e^-\to q\bar q q\bar q$ subprocess.
Things go as follows. Two-quark-two-gluon final states are produced
via the following eight Feynman graphs (here, 5 labels the 
$e^+e^-\to \gamma,Z$ current):
\begin{eqnarray}
 \nonumber
{T}_{1}=
\begin{picture}(150,50)
\SetScale{1.0}
\SetWidth{1.2}
\SetOffset(0,-60)
\Gluon(45,75)(30,90){3}{3}
\Gluon(30,40)(45,55){3}{3}
\Text(25,95)[]{\small 1}
\Text(25,35)[]{\small 2}
\ArrowLine(60,40)(45,55)
\ArrowLine(45,55)(45,75)
\ArrowLine(45,75)(60,90)
\DashLine(55,85)(70,85){2}
\Text(77.5,85)[]{\small 5}
\Text(65,35)[]{\small 4}
\Text(65,95)[]{\small 3}
\end{picture} 
{T}_{2}=
\begin{picture}(150,50)
\SetScale{1.0}
\SetWidth{1.2}
\SetOffset(0,-60)
\Gluon(45,75)(30,90){3}{3}
\Gluon(30,40)(45,55){3}{3}
\Text(25,95)[]{\small 1}
\Text(25,35)[]{\small 2}
\ArrowLine(60,40)(45,55)
\ArrowLine(45,55)(45,75)
\ArrowLine(45,75)(60,90)
\DashLine(45,65)(60,65){2}
\Text(67.5,65)[]{\small 5}
\Text(65,35)[]{\small 4}
\Text(65,95)[]{\small 3}
\end{picture} 
{T}_{3}=
\begin{picture}(150,50)
\SetScale{1.0}
\SetWidth{1.2}
\SetOffset(0,-60)
\Gluon(45,75)(30,90){3}{3}
\Gluon(30,40)(45,55){3}{3}
\Text(25,95)[]{\small 1}
\Text(25,35)[]{\small 2}
\ArrowLine(60,40)(45,55)
\ArrowLine(45,55)(45,75)
\ArrowLine(45,75)(60,90)
\DashLine(55,45)(70,45){2}
\Text(77.5,45)[]{\small 5}
\Text(65,35)[]{\small 4}
\Text(65,95)[]{\small 3}
\end{picture} 
\\ \nonumber
\\ \nonumber
\\ \nonumber
{T}_{4}=
\begin{picture}(150,50)
\SetScale{1.0}
\SetWidth{1.2}
\SetOffset(0,-60)
\Gluon(45,75)(30,90){3}{3}
\Gluon(30,40)(45,55){3}{3}
\Text(25,95)[]{\small 2}
\Text(25,35)[]{\small 1}
\ArrowLine(60,40)(45,55)
\ArrowLine(45,55)(45,75)
\ArrowLine(45,75)(60,90)
\DashLine(55,85)(70,85){2}
\Text(77.5,85)[]{\small 5}
\Text(65,35)[]{\small 4}
\Text(65,95)[]{\small 3}
\end{picture} 
{T}_{5}=
\begin{picture}(150,50)
\SetScale{1.0}
\SetWidth{1.2}
\SetOffset(0,-60)
\Gluon(45,75)(30,90){3}{3}
\Gluon(30,40)(45,55){3}{3}
\Text(25,95)[]{\small 2}
\Text(25,35)[]{\small 1}
\ArrowLine(60,40)(45,55)
\ArrowLine(45,55)(45,75)
\ArrowLine(45,75)(60,90)
\DashLine(45,65)(60,65){2}
\Text(67.5,65)[]{\small 5}
\Text(65,35)[]{\small 4}
\Text(65,95)[]{\small 3}
\end{picture} 
{T}_{6}=
\begin{picture}(150,50)
\SetScale{1.0}
\SetWidth{1.2}
\SetOffset(0,-60)
\Gluon(45,75)(30,90){3}{3}
\Gluon(30,40)(45,55){3}{3}
\Text(25,95)[]{\small 2}
\Text(25,35)[]{\small 1}
\ArrowLine(60,40)(45,55)
\ArrowLine(45,55)(45,75)
\ArrowLine(45,75)(60,90)
\DashLine(55,45)(70,45){2}
\Text(77.5,45)[]{\small 5}
\Text(65,35)[]{\small 4}
\Text(65,95)[]{\small 3}
\end{picture} 
\end{eqnarray}
\vskip-0.2cm
\begin{eqnarray}\label{QQgg}
\nonumber
\quad
\quad
\quad
\quad
{T}_{7}=
\begin{picture}(150,50)
\SetScale{1.0}
\SetWidth{1.2}
\SetOffset(0,-62.5)
\Gluon(15,50)(30,65){3}{3}
\Gluon(30,65)(15,80){3}{3}
\Text(10,45)[]{\small 2}
\Text(10,85)[]{\small 1}
\Gluon(30,65)(60,65){3}{3}
\ArrowLine(75,50)(60,65)
\ArrowLine(60,65)(75,80)
\DashLine(70,75)(85,75){2}
\Text(92.5,75)[]{\small 5}
\Text(80,45)[]{\small 4}
\Text(80,85)[]{\small 3}
\end{picture} 
{T}_{8}=
\begin{picture}(150,50)
\SetScale{1.0}
\SetWidth{1.2}
\SetOffset(0,-62.5)
\Gluon(15,50)(30,65){3}{3}
\Gluon(30,65)(15,80){3}{3}
\Text(10,45)[]{\small 2}
\Text(10,85)[]{\small 1}
\Gluon(30,65)(60,65){3}{3}
\ArrowLine(75,50)(60,65)
\ArrowLine(60,65)(75,80)
\DashLine(70,55)(85,55){2}
\Text(92.5,55)[]{\small 5}
\Text(80,45)[]{\small 4}
\Text(80,85)[]{\small 3}
\end{picture}
\\
\\ \nonumber
\end{eqnarray}

The two aforementioned colour flows are identified by the terms
proportional to
$(t^A t^B)_{i_3 i_4}$ and $(t^B t^A)_{i_3 i_4}$, where $A$ and $B$
are the colours of the gluons labelled $1$ and $2$, respectively.
In fact, it should be recalled that the colour pieces associated
with the triple-gluon vertices are nothing but the 
structure constants $f^{ABX}$ of the $SU(N_C)$ gauge group, for which
\begin{equation}\label{algebra}
[t^A,t^B]_{i_3i_4}\equiv (t^A t^B)_{i_3i_4}-(t^B t^A)_{i_3 i_4}={\mathrm{i}}
f^{ABX}t^X_{i_3i_4},
\end{equation}
$X$ being the label of the intermediate gluon 
and $i_3,i_4$ the outgoing quark
colour indices. Therefore, one can conveniently group the original eight  
Feynman amplitudes as 
\begin{equation}\label{M_ggQQPhi}
M_{+}=\sum_{i=1}^3 T_{i}
                     +\sum_{i=7}^8 T_{i},
\quad\quad\quad\quad\quad 
M_{-}=\sum_{i=4}^6 T_{i}
                     -\sum_{i=7}^8 T_{i}.
\end{equation}
The two 
positive definite contributions $N_{+}$ and $N_{-}$ corresponding to the two 
fundamental colour connections needed for the interface to the 
{\tt HERWIG} parton shower (see Ref.~\cite{PinoBryan}) can, for
example, be obtained as:  
$$
N_{+} =~D~     
      (|{M}_{+}|^2
      -\frac{1}{N_C^2}|{M}_{+}+{M}_{-}|^2
       \frac{|{M}_{+}|^2}
            {|{M}_{+}|^2+|{M}_{-}|^2}),
$$
\begin{equation}\label{hw}       
N_{-} =~D~    
      (|{M}_{-}|^2
      -\frac{1}{N_C^2}|{M}_{+}+{M}_{-}|^2
       \frac{|{M}_{-}|^2}
            {|{M}_{+}|^2+|{M}_{-}|^2}),
\end{equation}       
where the colour pre-factor is ($N_C=3$)
\begin{equation}
D=\frac{N_C}{4}(N_C^2-1),
\end{equation}
so that the total squared amplitude can be recovered from
\begin{equation}
{M}^2(q\bar q gg) = \sum_{i=\pm}{N}_i.
\end{equation}
In the right-hand sides of
eq.~(\ref{hw}), the first terms are
the `planar' whereas the second are the `non-planar' ones \cite{PinoBryan}.  
Here, $N_{+}$ would correspond to the $t$-channel colour flow 
$(4 \ar 2, 2 \ar 1, 1 \ar 3)$ and   $N_{-}$ to the $u$-channel one
$(4 \ar 1, 1 \ar 2, 2 \ar 3)$, that is, {\tt 2413} and {\tt 4123}
in the notation of Ref.~\cite{PinoBryan}. This is the default
procedure adopted in HERWIG 6.1 that we have maintained in our analysis
(see Ref.~\cite{herwig61} for other possible choices).

\begin{figure}[!t]
\vskip-2.0cm
~\hskip2.75cm\epsfig{file=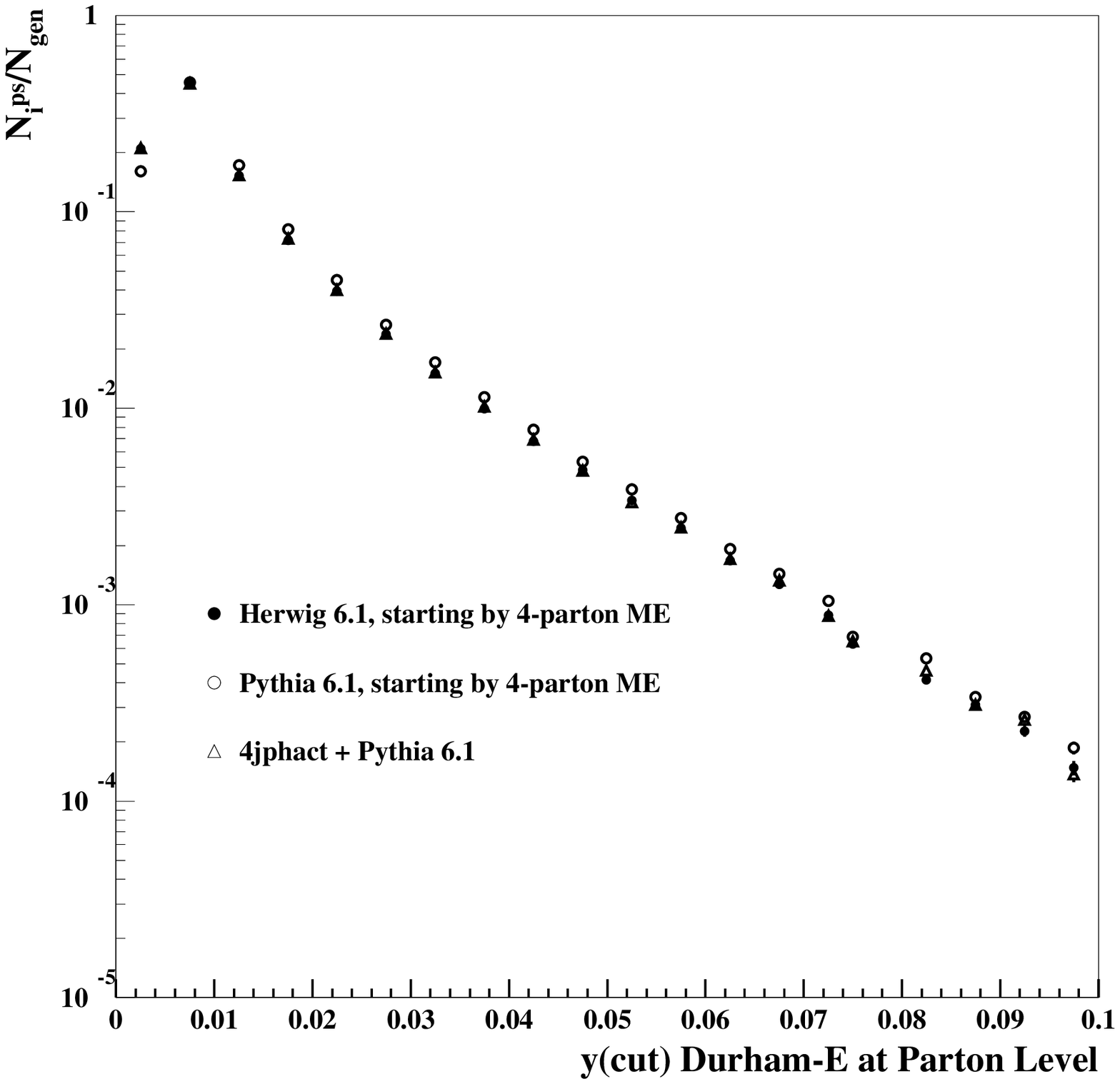,height=14cm,angle=0}
\vskip-4.40cm
~\hskip2.75cm\epsfig{file=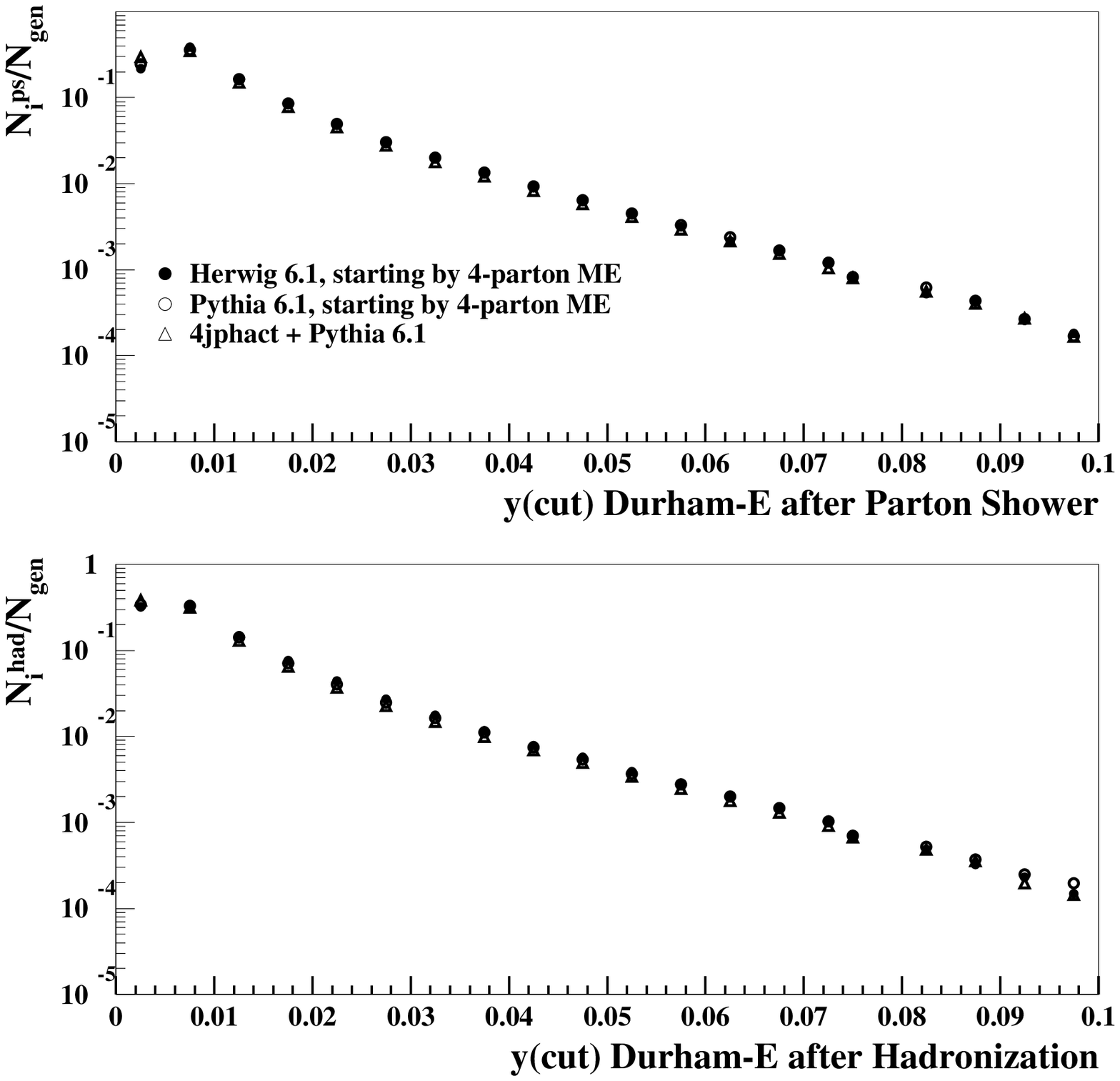,height=14cm,angle=0}
\vskip-2.50cm
\caption{\small The $y_{{cut}}$ dependence of four-jet events
at parton level before (top) and after (middle) showering effects
as well as at hadron level (bottom).}
\label{fig:y34}
\end{figure}

Several kinematic quantities can be used to disentangle Higgs
events from the backgrounds at a LC. One example is the
polar angle of the reconstructed $Z$ boson. In fact, 
in the case of the signal, one expects the gauge vector to be very central,
owning to the $s$-channel and spin dynamics of the Higgs process 
\cite{HiggsNLC}.
In contrast, no such a behaviour would appear in the QCD
noise. Fig.~\ref{fig:polar} shows this variable as obtained by using
${\cal O}(\alpha_s)  $ (from HERWIG 5.9, {\tt IPROC=105}) and 
${\cal O}(\alpha_s^2)$ (from HERWIG 6.1, {\tt IPROC=605}) MEs,
supplemented by the subsequent PS and the hadronisation stage.
The four-jet final state has been selected by using the Durham (D)
jet-clustering algorithm with resolution 
$y_{\mathrm{cut}}=0.001$\footnote{Notice that exactly 
four jets are required
to be reconstructed at hadron level. To allow for $n\ge4$ jets in the
final state (eventually clustered into four) would not alter
our conclusions. Note that we use the so-called E recombination
scheme: two tracks are clustered by
summing their four-momenta.}. Furthermore, we have imposed that one
pair of jets reproduce the $Z$ mass within 10 GeV, i.e.
$|M_{\mathrm{jj}}-M_Z|<10$ GeV, as illustration of a typical 
selection procedure of $e^+e^-\to HZ\to 4~\mathrm{jet}$ events.
These are the two jets that we will consider to be produced in
the $Z\to$~2~jet decay. 
For reference, we also have superimposed the same angular spectrum
as obtained from the Higgs process, again
generated by using HERWIG ({\tt IPROC=305}).
As self-evident from the choice of the {\tt IPROC} numbers, we
always require two $b$-jets in the final state, for both QCD and
Higgs processes.
The HERWIG default choice $M_H=150$ GeV is used for the SM Higgs mass.
Also notice that we normalise the cross sections
of all processes to unity. In other terms, we implicitly assume that
the two QCD descriptions produce the same event rate, for a given choice
of $\alpha_s$, jet-clustering scheme and $y_{\mathrm{cut}}$.
In reality, there exist differences in the total production rates
between the two QCD implementations, that could be source
of further systematic uncertainties. However, we leave these aside for the
time being, as we only concentrate on the kinematic effects.
It is clear from Fig.~\ref{fig:polar} that not only the two QCD
descriptions show significant differences between each other
in the polar angle distribution of the di-jet system emulating
a $Z$ decay, but these also persist where the Higgs process accumulates.
Similar effects can be appreciated in the distribution
of the relative angle between the two jets reconstructing the gauge
vector, see Fig.~\ref{fig:cjj}. This quantity too is used in the
selection of Higgs events at a LC \cite{HiggsNLC}.
Thus, although the behaviour of the two QCD implementations is rather
similar in the vicinity of the Higgs peak, see Fig.~\ref{fig:MH},
the effect of cuts enforced on $\cos\theta_{\mathrm{beam}}$
and/or $\cos\theta_{\mathrm{jj}}$ can be dramatically
different for the actual number of background events falling 
around $M_{\mathrm{recoil}}\approx M_H$. This can be appreciated
in Tab.~\ref{tab}.

\begin{table}[!h]
\begin{center}
\begin{tabular}{|c|c|c|}

\hline
$HZ$ & ${\cal O}(\alpha_s)$ & ${\cal O}(\alpha_s^2)$  \\
\hline\hline
$72.~\%$ & $9.8~\%$ & $12.~\%$ \\
\hline
\end{tabular}
\end{center}
\caption{Percentage of four-jet events in a window 
$|M_{\mathrm{recoil}}-M_H|<30$ GeV, after the cuts 
$|\cos\theta_{\mathrm{beam}}|<0.6$ and $\cos\theta_{\mathrm{jj}}<0.5$.}
\label{tab}
\end{table}

As mentioned earlier on, at least three other MCs exist that
have the option of generating multi-jet final states starting
from the ${\cal O}(\alpha_s^2)$ MEs. Among these, we consider
here PYTHIA 6.1 and 4JPHACT. The former makes use of the same (massless)
MEs implemented in HERWIG (with a similar approximate
procedure to account for mass effects \cite{Andre}), the main differences 
being in the treatment
of the PS and hadronisation. The latter, while using
showering and fragmentation dynamics borrowed from PYTHIA, employs
the fully massive MEs of Ref.~\cite{noi} and a massive four-body
phase space as well. 
A comparison among these MCs has been performed by Silvia Bravo for ALEPH
and we present here just a selection of the results she 
obtained\footnote{Others can be found at
{\tt http://www.pd.infn.it/ecfa/.}}. 
This is done in Fig.~\ref{fig:y34}, where the
$y_{\mathrm{cut}}$ dependence of the four-jet rates
 (all flavour combinations), as obtained
in the three MC environments, is presented at three
different stages. First, at the four-parton level before
showering effects, then after the latter have taken place, finally
after hadronisation. The comparison is performed at $\sqrt s=M_Z$ and
exactly four jets are required to be selected by the Durham
jet finder (using E-scheme recombination), for a given
$y_{\mathrm{cut}}$ value. Normalisation is to the number of generated
events: 1,000,000 in each case. The overall agreement is quite good,
particularly at the experimentally more relevant hadron level. In the end,
systematic errors induced by the different treatment of MEs, phase space,
PS and hadronisation in the three MCs are typically much smaller
than those between any ${\cal O}(\alpha_s^2)$
description and the standard ${\cal O}(\alpha_s)$ ones.

In summary, we have shown that there exist sizable differences
in the MC simulation of key LC phenomenology in four-jet final states, 
depending upon whether $e^+e^-\to q\bar q g$ or
$e^+e^-\to q\bar qgg$ plus $e^+e^-\to q\bar q Q\bar Q$ partonic
scatterings are used to initiate the PS of the event generator. However, 
several reliable
MC programs exist by now, each exploiting a different implementation 
of either approach.
Thus, new, more sophisticated analyses of four-jet events
will soon be available, these allowing one to pin down to
even better accuracy the physic potential of a future $e^+e^-$ collider
operating at the near-TeV scale.

Before concluding, we would like to briefly remark on other  aspects
that enter the description of hadronic final states with arbitrary jet
multiplicity, 
that were beyond the scope of this note. Here, we have been concerned
with  what 
we should like to call the procedure of `interfacing' MEs to the
subsequent PS, that is, of generating hadronic events through higher-order MEs,
but limiting the event generation to a phase space
region where the former are reliable, without any attempt to resort to
the latter otherwise. In fact, another
important issue is that of `matching' MEs and PS, by which
we mean the procedure of exploiting a combined approach that uses
MEs for large phase space separations, PS in the infrared limit
and modified MEs (by Sudakov form factors) in the overlapping region
\cite{Bryan}.
This is what is currently done within the APACIC++ event generator
\cite{apacic}, which uses exact leading-order (LO) MEs for
$e^+e^-\to n~{\mathrm{partons}}$, with $n\le5$\footnote{A similar
implementation, for $n\le4$, is also provided 
in the HERWIG  context \cite{Bryanweb}.}.
Note that the procedure of Ref.~\cite{Bryan}
is correct to next-to-leading logarithmic (NLL) 
accuracy and could possibly be extended to next-to-leading order (NLO) as well
\cite{Collins}. The interface/matching to the PS
 of higher-order MEs including loops \cite{ERT,a3}
(hence, generating negative weights) remains instead 
a challenge for the future \cite{potter}. 
\vskip0.15cm\noindent
\underbar{\sl Acknowledgements:}~I would like to thank Silvia Bravo
for the kind permission to reproduce here some of her unpublished material.
Useful conversation with A. Ballestrero, M.H. Seymour, T. Sj\"{o}strand and 
B.R. Webber are also acknowledged.


\begin{thebibliography}{99}
\def\baselinestretch{0.75}
{\small

\bibitem{herwig61} G. Corcella, I.G. Knowles, G. Marchesini, S. Moretti, 
K. Odagiri, P. Richardson, M.H. Seymour and B.R. Webber,
preprint Cavendish-HEP-99/17, December 1999,
{\tt hep-ph/9912396}.

\bibitem{HERWIG} G.\ Marchesini, B.R.\ Webber, G.\ Abbiendi, I.G.\ Knowles,
  M.H.\ Seymour and L.\ Stanco, \cpc 67 1992 465.

\bibitem{ee500} See for example:
Proceedings of the Workshop `$e^+e^-$ Collisions at
500 GeV. The Physics Potential', 
Munich, Annecy, Hamburg, 3--4 February 1991, ed. P.M.~Zerwas, DESY 92--123A/B,
August 1992, DESY 93--123C, December 1993.


\bibitem{HiggsNLC} See for example:
`Higgs Particles' sections, in Ref.~\cite{ee500}.

\bibitem{Bjorken} J.D.~Bjorken, Proceedings of the 
                 `{ Summer Institute on Particle
                 Physics}', {\it SLAC Report} 198 (1976);
                 B.W.~Lee, C.~Quigg and H.B.~Thacker, {\it Phys. Rev}
                 {\bf D16} (1977) 1519;
                 B.L.~Ioffe and V.A.~Khoze, 
                 {\it Sov. J. Part. Nucl.} {\bf 9} (1978) 50;
                 J.~Ellis, M.K.~Gaillard and D.V.~Nanopoulos, 
                 \np B106 1976 292.

\bibitem{ALEPHgluino} ALEPH Collaboration, 
{\it Z. Phys.}
{\bf C76} (1997) 1.

\bibitem{ioebas} S.~Bethke, A.~Ricker and P.M.~Zerwas,
{\it Z.~Phys.} {\bf C49} (1991) 59;
S. Moretti and J.B. Tausk, {\it Z. Phys.} {\bf C69} (1996)
635.

\bibitem{wrksp} A. Ballestrero {\it et al.}, {\it J. Phys.} 
{\bf G24} (1998) 365.

\bibitem{GC} G. Cowan, {\it J. Phys.} {\bf G24} (1998) 307.

\bibitem{Ian} I.G. Knowles, {\it Nucl. Phys.} {\bf B310} (1988) 571,
{\it J. Phys.} {\bf G17} (1991) 1562.
 
\bibitem{OP} OPAL Collaboration, \zp C65 1995 367. 

\bibitem{Andre} J. Andr\'e and T.\ Sj\"ostrand, 
{\it Phys. Rev.} {\bf D57} (1998) 5767;
J. Andr\'e, \preprint\ LU-TP-97-12, June 1997, \hepph 9706325.

\bibitem{pythia61} T.\ Sj\"ostrand, {\it Comp. Phys. Commun.}
{\bf 39} (1984) 347;
                 M.\ Bengtsson and T.\ Sj\"ostrand, 
 {\it Comp. Phys. Commun.} {\bf 43} (1987) 367;
for the latest PYTHIA version, see:\\
{\tt http://www.thep.lu.se/staff/torbjorn/Pythia.html}.

\bibitem{4jphact} A. Ballestrero, in preparation.

\bibitem{apacic} F. Krauss, R. Kuhn and G. Soff,
{\it J. Phys.} {\bf G26} (2000) L11, {\it Acta Phys. Polon.}
{\bf B30} (1999) 3875.

\bibitem{spin} S. Moretti and W.J. Stirling,
{\it Eur. Phys. J.} {\bf C9} (1999) 81.

\bibitem{ERT} R.K. Ellis, D.A. Ross and A.E. Terrano,
{\it Nucl. Phys.} {\bf B178} (1981) 421.

\bibitem{CataniSeymour}
S. Catani and M.H. Seymour, {\it Phys. Lett.} {\bf B378} (1996) 287;
see also:\\ 
{\tt http://hepwww.rl.ac.uk/theory/seymour/nlo/.}

\bibitem{schemes} S. Moretti, L. L\"onnblad, T. Sj\"ostrand,
{\it JHEP} {\bf 08} (1998) 001.

\bibitem{herwig62} G. Corcella, I.G. Knowles, G. Marchesini, S. Moretti, 
K. Odagiri, P. Richardson, M.H. Seymour and B.R. Webber, 
in preparation.

\bibitem{kosuke} K. Odagiri, {\it JHEP} {\bf 10} (1998) 006.

\bibitem{PinoBryan} G.~Marchesini and B.R.~Webber, {\it Nucl.~Phys.}~{\bf B310}
(1988) 461. 

\bibitem{noi} A. Ballestrero, E. Maina and S. Moretti,
           {\it Phys. Lett.} {\bf B294} (1992) 425,
           {\it Nucl. Phys.} {\bf B415} (1994) 265,
      Proceedings of the `XXIXth Rencontres de Moriond', 
     M\'eribel, France,
      March 1994, ed. J. Tr\^an Thanh V\^an (ed.
      Fronti\`eres, Gif-sur-Yvette, 1994), page  367.

\bibitem{Bryan} B.R. Webber, 
 Talk given at `XXXVth Rencontres de Moriond', Les Arcs, France, March 2000,
preprint Cavendish-HEP-00/05, May 2000, {\tt hep-ph/0005035}.

\bibitem{Bryanweb} See: {\tt http://webber.home.cern.ch/webber/.}

\bibitem{Collins} J. Collins, {\it JHEP} {\bf 0005} (2000) 004.

\bibitem{a3} Z.~Bern, L.~Dixon, D.A.~Kosower and S.~Weinzierl,
{\it Nucl. Phys.} {\bf B489} ({1997}) {3};
Z.~Bern, L.~Dixon, D.A.~Kosower, 
{\it Nucl. Phys.} {\bf B513} (1998) 3;
L.~Dixon and  A.~Signer, {\it Phys. Rev. Lett.} {\bf 78} ({1997}) {811};
E.W.N.~Glover and D.J.~Miller, {\it Phys.~Lett.} {\bf B396} ({1997}) {257};
J.M.~Campbell, E.W.N.~Glover and D.J.~Miller,
{\it Phys.~Lett.} {\bf B409} (1997) 503.

\bibitem{potter} B. P\"otter, preprint MPI/PhT/2000-24, July 2000, {\tt hep-ph/0007172}.
}
\end{thebibliography}
\end{document}